\documentclass[a4paper,11pt]{article}

\usepackage{inputenc}
\usepackage[T1]{fontenc}
\usepackage[english]{babel}

\usepackage{indentfirst}
\usepackage{amsmath}
\usepackage{amssymb}
\usepackage{amsthm}
\usepackage{graphicx}
\usepackage{subfigure}
\usepackage{psfrag}{}
\usepackage{color}
\usepackage{pgf}

\allowhyphens

\newcommand{\fa}{\mathfrak{a}}

\newcommand{\valos}{\mathbb{R}}
\newcommand{\complex}{\mathbb{C}}

\newcommand{\eps}{\varepsilon}

\newcommand{\ordo}{\mathcal{O}}

\newcommand{\ket}[1]{{\left|#1\right\rangle}}
\newcommand{\bra}[1]{{\left\langle #1\right|}}

\newcommand{\skalarszorzat}[2]{{\langle #1 | #2 \rangle}}


\makeatother

\begin{document}

\numberwithin{equation}{section}

\title{Dynamical free energy and the Loschmidt-echo for a class of quantum quenches
  in the Heisenberg spin chain}
\author{Bal\'azs Pozsgay$^1$\\
~\\
 $^{1}$MTA-BME \textquotedbl{}Momentum\textquotedbl{} Statistical
Field Theory Research Group\\
1111 Budapest, Budafoki \'ut 8, Hungary
}

\maketitle

\abstract{We consider a class of global quantum quenches in the Heisenberg XXZ
  spin chain, where the initial states are given by products of local
  two-site states. 
The two main examples are the N\'eel state and the dimer state. We
derive an exact analytic result for the ,,Loschmidt echo per site'' at
imaginary times and also consider the analytic continuation back to
real times. As a by-product we obtain an exact result for the
,,overlap per site'' between the N\'eel state and the ground state of
the XXZ Hamiltonian in the massive regime.
}
\section{Introduction}

Non-equilibrium dynamics of quantum systems, in particular quantum
quenches attracted a lot of interest lately. One of the main questions
is whether an isolated quantum system equilibrates, and if yes, what
are the stationary values of physical observables and how do they
depend on the initial state and the Hamiltonian governing the time
evolution. Generally one expects thermalization, which means that
 the stationary values of observables coincide with those
calculated from a thermal ensemble with a given temperature
\cite{Silva-quench-colloquium}. 

Integrable models provide an interesting setting where the real-time
dynamics can be markedly different. These theories possess higher
conserved charges, which prevent thermalization in the usual
sense. Instead, it was proposed in \cite{rigol-gge} that in the long-time
limit the mean values of observables are given by 
 the so-called Generalized Gibbs Ensemble (GGE), which incorporates all
 conserved charges with appropriate Lagrange-multipliers.

Most previous work focused on integrable spin chains which are
equivalent to free fermions
\cite{mussardo-ising-1,mussardo-ising-2,ising-quench-1,ising-quench-2,ising-quench-3,ising-quench-4,ising-quench-5,free-gge-1,free-gge-2,free-gge-3,free-gge-4,essler-truncated-gge}
or models which can be treated by Conformal Field Theory methods
\cite{cardy-calabrese}.
Considerable work has been spent to understand the quench dynamics of
the 1D Bose gas, especially in the infinite interaction limit
\cite{marci-ll-quench1,marci-ll-quench2,spyros2}. 
 Concerning interacting theories not equivalent to free fermions
there are only few analytic results available \cite{2013PhRvA..87e3628I,spyros1} and even the
numerical treatment based on the Bethe Ansatz solution is challenging \cite{caux-brandino}.

The Heisenberg spin chain is a paradigmatic integrable model, which is
one of the simplest genuinely interacting models. Yet, the questions
of equilibration and thermalization are very far from being
solved. 
The first two papers to consider the GGE for the XXZ spin chain were
\cite{sajat-xxz-gge} and \cite{essler-xxz-gge}. Both developed the
so-called Quantum Transfer Matrix formalism for the GGE and gave
approximate predictions for the long-time limit of local correlators
following a quench from the N\'eel state.
In \cite{sajat-xxz-gge} a truncated GGE was used, whereas
\cite{essler-xxz-gge} took into account all higher charges in a
$1/\Delta$ expansion. We would like to stress that both papers bypass
the derivation of the actual time dependence and they \textit{assume}
the GGE hypothesis to provide predictions which could be checked by
other methods.

One of the reasons for the lack of rigorous results for
non-equilibrium dynamics in the XXZ spin chain is that
the celebrated Bethe Ansatz solution is only adequate to equilibrium
problems. Calculation of the real time dynamics requires to take into
account all states (or at least a large subset of them
\cite{quench-action}), which is a notoriously difficult
task. Moreover, one needs manageable formulas for overlaps of Bethe
states and the initial state, which are typically not available either.

Apart from the correlation functions there has been recent interest
 in
the Loschmidt echo (the overlap of the initial and time-evolved
states) too, which is a somewhat simpler quantity, yet it possesses
unexpected features. In \cite{gs-Loschmidt-1,gs-Loschmidt-2} it was
observed that the Loschmidt echo displays non-analytic behaviour if
the system is quenched across a critical point. This problem was
considered in the  very recent work \cite{Fagotti-Loschmidt},
where M. Fagotti derived analytic results for the Loschmidt echo per
site in the XXZ
chain using the GGE
hypothesis for the exponential of the Hamiltonian. 
The non-linear integral equations (NLIE) of \cite{Fagotti-Loschmidt} are
implicit in the sense that Lagrange-multipliers are not specified,
nevertheless it was shown that numerical results can be calculated in the
$1/\Delta$ approximation and the analytic properties of functions
involved in the NLIE can be studied.

In the present work we derive an exact analytic result for the 
Loschmidt echo per site for the case of purely imaginary times and we also
consider the analytic continuation back to real times.  To our best
knowledge this is the first exact result concerning the
time-dependence of a physical quantity in a non-equilibrium
setting of the interacting XXZ chain.

The structure of the paper is as follows. In Section 2 we present
the problem and the general notations. In Section 3 we develop the
Trotter-Suzuki decomposition for the Loschmidt echo at imaginary times
and show that the problem is equivalent to finding the leading eigenvalue of
the so-called Boundary Quantum Transfer Matrix (BQTM). In Section 4 we
diagonalize the BQTM using the already available techniques of the
Boundary Algebraic Bethe Ansatz. In Section 5 the resulting equations
are worked out for a quench from the N\'eel state. Subsection 5.4
includes numerical results and our investigations about the analytic
continuation to real times. Section 6 is devoted to the XXX limit and
the quench starting from the dimer  state. We conclude in
Section 7.

\section{Dynamical free energy density and the Loschmidt echo}

Consider a quantum quench situation in a 1D spin chain, where at $t=0$ the system is prepared in a state $\ket{\Psi_0}$ and for $t>0$ its time evolution is governed by a Hamiltonian of the form
\begin{equation}
\label{Hgen}
H=\sum_{j=1}^L  u_j+h_{j,j+1},
\end{equation}
where $u_j$ and $h_{j,j+1}$ are one-site and two-site operators, $L$ is the length of the chain and periodic boundary conditions are assumed. The state $\ket{\Psi_0}$ can be chosen as ground state of an other local Hamiltonian $H_0$, or it can be prepared according to some well-defined rule. The specific examples for $\ket{\Psi_0}$ will be given later.

In the present work we focus on the cumulant generating function and the Loschmidt echo. 
The cumulant generating function $G(s)$ is defined as
\begin{equation}
\label{GGs}
  \exp{G(s)}= \bra{\Psi_0}\exp(-sH)\ket{\Psi_0},\quad \quad s\in \valos^+.
\end{equation}
It satisfies the initial condition $G(0)=0$ and its
power series in $s$ 
 defines the cumulants of
$H$:
\begin{equation}
  G(s)=\sum_{n=1}^\infty \kappa_n \frac{s^n}{n!}.
\end{equation}
The first few cumulants are
\begin{equation}
  \begin{split}
   \kappa_1&=-\bra{\Psi_0}H\ket{\Psi_0}\\
  \kappa_2&=\bra{\Psi_0}H^2\ket{\Psi_0}-\bra{\Psi_0}H\ket{\Psi_0}^2\\
\kappa_3&=-\bra{\Psi_0}H^3\ket{\Psi_0}+3\bra{\Psi_0}H^2\ket{\Psi_0}\bra{\Psi_0}H\ket{\Psi_0}-2\bra{\Psi_0}H\ket{\Psi_0}^3.
  \end{split}
\end{equation}
It is easy to see that every $\kappa_n$ is linear in $L$ once $L>n$. 

It is useful to introduce the function 
\begin{equation}
\label{gsdef}
  g(s)=\lim_{L\to\infty}\frac{G(s)}{L}.
\end{equation}
Its power series expansion is given by
\begin{equation}
\label{gs}
  g(s)=\sum_{n=1}^\infty \tilde \kappa_n \frac{s^n}{n!}
\end{equation}
with 
\begin{equation*}
  \tilde \kappa_n=\lim_{L\to\infty} \frac{\kappa_n}{L}.
\end{equation*}
Note that even though $g(s)$ is well defined through \eqref{gsdef}, the
expansion \eqref{gs} is typically not a convergent series, as the $L\to\infty$ limit is not uniformly convergent.

The function $g(s)$ is the main object of interest of this work. It is analogous to the free energy density if $s$ is interpreted as an inverse temperature. Following \cite{Fagotti-Loschmidt} we call it the dynamical free energy density. 

The Loschmidt echo is defined as the overlap of the time evolved state and the initial state:
\begin{equation*}
  M(t)=\left|\bra{\Psi_0}\exp(-itH)\ket{\Psi_0}\right|^2.
\end{equation*}
For any finite $t$ it decays exponentially with the volume. Therefore it is useful to define the ``Loschmidt echo per site'':
\begin{equation*}
  m(t)=M(t)^{1/L}.
\end{equation*}
It follows from the formulas above that the Loschmidt echo per site is given by the analytic continuation of the dynamical free energy density:
\begin{equation*}
 \log m(t)=2\Re(g(it)).
\end{equation*}

At any real $s$ the object in \eqref{GGs} is equivalent to a partition function of a 2D classical system with boundary conditions specified by the state $\ket{\Psi_0}$. 
This correspondence will be used below to derive exact analytic results for $g(s)$, $s\in\valos^+$ for a class of quantum quenches in the XXZ Heisenberg spin chain.
The analytic continuation to $s=it$ is considered in subsection \ref{sec:Loschmidt}.

\section{Hamiltonian, initial states, and the Trotter-Suzuki decomposition}

The Hamiltonian of the XXZ Heisenberg spin chain is 
\begin{equation}
  \label{XXZ-H}
  H=\sum_{j=1}^{L}
  (\sigma^x_j\sigma^x_{j+1}+\sigma^y_j\sigma^y_{j+1}+\Delta
(\sigma^z_j\sigma^z_{j+1}-1)).
\end{equation}
The anisotropy parameter $\Delta$ can be chosen arbitrarily, the $\Delta=1$ case specifies the $SU(2)$ symmetric XXX Hamiltonian.

We consider initial states $\ket{\Psi_0}$ which are constructed as products of two-site states:
\begin{equation}
\label{psi0}
  \ket{\Psi_0}=\ket{v}\otimes \ket{v}\otimes \dots \otimes \ket{v}.
\end{equation}
Here $\ket{v}\in \complex^2\otimes \complex^2$ is of the form
\begin{equation}
\label{vdef}
  \ket{v}=\frac{1}{\sqrt{1+|\gamma|^2}} 
(\ket{+-}+\gamma \ket{-+}).
\end{equation}
The following two particular cases will be considered:
\begin{itemize}
\item The case $\gamma=0$ which is simply the N\'eel state
  $\ket{N}=\ket{+-+-\dots}$. This state is one of the two ground
  states of the Hamiltonian in the $\Delta\to\infty$ limit. 
\item The case $\gamma=-1$ which we call the fully dimerized, or
  simply dimer state. This state is $SU(2)$ symmetric and it is
one of the two ground  states of the Majumdar-Ghosh Hamiltonian
\cite{MG}
\begin{equation*}
  H_{MG}=\sum_j \sigma_j\cdot \sigma_{j+1}+\frac{1}{2}\sum_j\sigma_j\cdot \sigma_{j+2}.
\end{equation*}
\end{itemize}
In both cases the other ground state is obtained by translation
by one site.

\subsection{Suzuki-Trotter decomposition}

The expectation value in \eqref{GGs} can be computed using the
Suzuki-Trotter decomposition
\begin{equation}
\exp(-sH)=\lim_{N\to\infty}
\left(1-\frac{sH}{N}  \right)^N.
\end{equation}
In order to evaluate the $N$-fold product we use the Algebraic Bethe Ansatz \cite{korepinBook} and the Quantum Transfer Matrix technique \cite{kluemper-QTM,kluemper-review}. As a first step we introduce the (rapidity dependent) monodromy matrix of the periodic spin chain as
\begin{equation}
\label{T}
  T(u)=\mathcal{L}_L(u)\dots \mathcal{L}_1(u).
\end{equation}
Here $\mathcal{L}_j(u)$ are local Lax-operators given by
\begin{equation}
\label{Lj}
  \mathcal{L}_j(u)=R_{0j}(u),
\end{equation}
where $R(u)$ is the trigonometric R-matrix given by
\begin{equation}
  R(u)=
  \begin{pmatrix}
    \sinh(u+\eta) & & &\\
& \sinh(u)  & \sinh(\eta) & \\
& \sinh(\eta) & \sinh(u) & \\
& & & \sinh(u+\eta)
  \end{pmatrix},
\label{R}
\end{equation}
where the parameter $\eta$ is related to the anisotropy by $\Delta=\cosh\eta$. The index
$j$ in \eqref{Lj} refers to the 
site $j$ of the spin chain, whereas $0$ refers to the so-called
auxiliary space. The monodromy matrix is represented pictorially in
figure \ref{fig:mono}. 

\begin{figure}
\centering
\begin{pgfpicture}{0cm}{-0.2cm}{14cm}{2.2cm}
\pgfsetendarrow{\pgfarrowto}

\pgfline{\pgfxy(1.5,1)}{\pgfxy(0.5,1)}  
\pgfline{\pgfxy(1,0.5)}{\pgfxy(1,1.5)}

\pgfputat{\pgfxy(1.8,1)}{\pgfbox[center,center]{$+$}}
\pgfputat{\pgfxy(1,0.2)}{\pgfbox[center,center]{$+$}}
\pgfputat{\pgfxy(1,1.8)}{\pgfbox[center,center]{$+$}}
\pgfputat{\pgfxy(0.2,1)}{\pgfbox[center,center]{$+$}}

\pgfputat{\pgfxy(3.3,1)}{\pgfbox[center,center]{$\sinh(u+\eta)$}}

\pgfline{\pgfxy(6.6,1)}{\pgfxy(5.6,1)}  
\pgfline{\pgfxy(6.1,0.5)}{\pgfxy(6.1,1.5)}

\pgfputat{\pgfxy(6.9,1)}{\pgfbox[center,center]{$+$}}
\pgfputat{\pgfxy(6.1,0.2)}{\pgfbox[center,center]{$-$}}
\pgfputat{\pgfxy(6.1,1.8)}{\pgfbox[center,center]{$-$}}
\pgfputat{\pgfxy(5.3,1)}{\pgfbox[center,center]{$+$}}

\pgfputat{\pgfxy(7.4,1)}{\pgfbox[left,center]{$\sinh(u)$}}

\pgfline{\pgfxy(11.5,1)}{\pgfxy(10.5,1)}  
\pgfline{\pgfxy(11,0.5)}{\pgfxy(11,1.5)}

\pgfputat{\pgfxy(11.8,1)}{\pgfbox[center,center]{$+$}}
\pgfputat{\pgfxy(11,0.2)}{\pgfbox[center,center]{$-$}}
\pgfputat{\pgfxy(11,1.8)}{\pgfbox[center,center]{$+$}}
\pgfputat{\pgfxy(10.2,1)}{\pgfbox[center,center]{$-$}}

\pgfputat{\pgfxy(12.3,1)}{\pgfbox[left,center]{$\sinh(\eta)$}}
\end{pgfpicture}
\caption{The 6-vertex model weights as specified by the trigonometric $R$-matrix \eqref{R}. Here $u$ is attached to the horizontal line and $0$ to the vertical
one.}
\label{fig:6}
\end{figure}

\begin{figure}
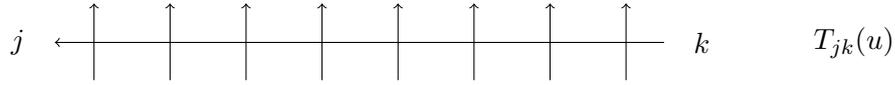

\centering
\begin{pgfpicture}{0cm}{-0.2cm}{11cm}{2.2cm}
\pgfsetendarrow{\pgfarrowto}

\pgfline{\pgfxy(8.5,1)}{\pgfxy(0.5,1)}  
\pgfline{\pgfxy(1,0.5)}{\pgfxy(1,1.5)}
\pgfline{\pgfxy(2,0.5)}{\pgfxy(2,1.5)}
\pgfline{\pgfxy(3,0.5)}{\pgfxy(3,1.5)}
\pgfline{\pgfxy(4,0.5)}{\pgfxy(4,1.5)}
\pgfline{\pgfxy(5,0.5)}{\pgfxy(5,1.5)}
\pgfline{\pgfxy(6,0.5)}{\pgfxy(6,1.5)}
\pgfline{\pgfxy(7,0.5)}{\pgfxy(7,1.5)}
\pgfline{\pgfxy(8,0.5)}{\pgfxy(8,1.5)}

\pgfputat{\pgfxy(0,1)}{\pgfbox[center,center]{$j$}}
\pgfputat{\pgfxy(9,1)}{\pgfbox[center,center]{$k$}}
\pgfputat{\pgfxy(11,1)}{\pgfbox[center,center]{$T_{jk}(u)$}}


\end{pgfpicture}
\caption{The monodromy matrix of the periodic system of length
  $L$. The horizontal line carries rapidity $u$, whereas the
  inhomogeneities of the vertical lines are zero. The $R$-matrix acts
  on the vertices with matrix elements given by the 6-vertex model
  weights depicted on figure \ref{fig:6}. }
\label{fig:mono}
\end{figure}

\begin{figure}
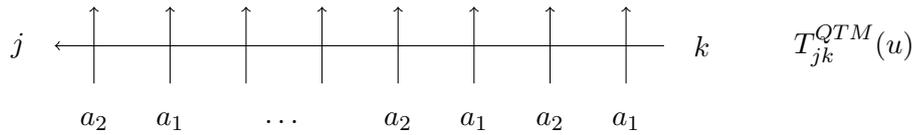

\centering
\begin{pgfpicture}{0cm}{-0.2cm}{11cm}{2.2cm}
\pgfsetendarrow{\pgfarrowto}

\pgfline{\pgfxy(8.5,1)}{\pgfxy(0.5,1)}  
\pgfline{\pgfxy(1,0.5)}{\pgfxy(1,1.5)}
\pgfline{\pgfxy(2,0.5)}{\pgfxy(2,1.5)}
\pgfline{\pgfxy(3,0.5)}{\pgfxy(3,1.5)}
\pgfline{\pgfxy(4,0.5)}{\pgfxy(4,1.5)}
\pgfline{\pgfxy(5,0.5)}{\pgfxy(5,1.5)}
\pgfline{\pgfxy(6,0.5)}{\pgfxy(6,1.5)}
\pgfline{\pgfxy(7,0.5)}{\pgfxy(7,1.5)}
\pgfline{\pgfxy(8,0.5)}{\pgfxy(8,1.5)}

\pgfputat{\pgfxy(0,1)}{\pgfbox[center,center]{$j$}}
\pgfputat{\pgfxy(9,1)}{\pgfbox[center,center]{$k$}}
\pgfputat{\pgfxy(11,1)}{\pgfbox[center,center]{$T_{jk}^{QTM}(u)$}}

\pgfputat{\pgfxy(8,0)}{\pgfbox[center,center]{$a_1$}}
\pgfputat{\pgfxy(7,0)}{\pgfbox[center,center]{$a_2$}}
\pgfputat{\pgfxy(6,0)}{\pgfbox[center,center]{$a_1$}}
\pgfputat{\pgfxy(5,0)}{\pgfbox[center,center]{$a_2$}}
\pgfputat{\pgfxy(3.5,0)}{\pgfbox[center,center]{$\dots$}}
\pgfputat{\pgfxy(2,0)}{\pgfbox[center,center]{$a_1$}}
\pgfputat{\pgfxy(1,0)}{\pgfbox[center,center]{$a_2$}}

\end{pgfpicture}
\caption{The so-called quantum monodromy matrix, which acts on an auxiliary spin chain of length $2N$, where $N$ is the Trotter-number. The inhomogeneities
associated to the vertical lines are $a_1=\beta/2N$ and $a_2=-\beta/2N-\eta$.}
\end{figure}

\begin{figure}
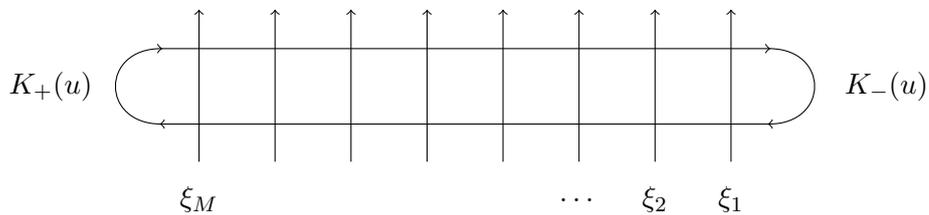

\centering
\begin{pgfpicture}{0cm}{4cm}{12cm}{7cm}

\pgfsetendarrow{\pgfarrowto}
\pgfsetstartarrow{\pgfarrowswap{\pgfarrowto}}

\pgfline{\pgfxy(9.5,5)}{\pgfxy(1.5,5)}
\pgfline{\pgfxy(1.5,6)}{\pgfxy(9.5,6)}

\pgfclearstartarrow

\pgfputat{\pgfxy(10.5,5.5)}{\pgfbox[left,center]{$K_-(u)$}}
\pgfputat{\pgfxy(-0.5,5.5)}{\pgfbox[left,center]{$K_+(u)$}}

\pgfputat{\pgfxy(9,4)}{\pgfbox[center,center]{$\xi_1$}}
\pgfputat{\pgfxy(8,4)}{\pgfbox[center,center]{$\xi_2$}}
\pgfputat{\pgfxy(7,4)}{\pgfbox[center,center]{$\dots$}}
\pgfputat{\pgfxy(2,4)}{\pgfbox[center,center]{$\xi_M$}}

\pgfline{\pgfxy(2,4.5)}{\pgfxy(2,6.5)}
\pgfline{\pgfxy(3,4.5)}{\pgfxy(3,6.5)}
\pgfline{\pgfxy(4,4.5)}{\pgfxy(4,6.5)}
\pgfline{\pgfxy(5,4.5)}{\pgfxy(5,6.5)}
\pgfline{\pgfxy(6,4.5)}{\pgfxy(6,6.5)}
\pgfline{\pgfxy(7,4.5)}{\pgfxy(7,6.5)}
\pgfline{\pgfxy(8,4.5)}{\pgfxy(8,6.5)}
\pgfline{\pgfxy(9,4.5)}{\pgfxy(9,6.5)}

\pgfclearendarrow

\pgfmoveto{\pgfxy(9.5,5)}
\pgfcurveto{\pgfxy(10.3,5)}{\pgfxy(10.3,6)}{\pgfxy(9.5,6)}
\pgfstroke

\pgfmoveto{\pgfxy(1.5,5)}
\pgfcurveto{\pgfxy(0.7,5)}{\pgfxy(0.7,6)}{\pgfxy(1.5,6)}
\pgfstroke

\end{pgfpicture}
\caption{The boundary transfer matrix with inhomogeneities $\xi_j$.}
\label{startingpoint}
\end{figure}

\begin{figure}
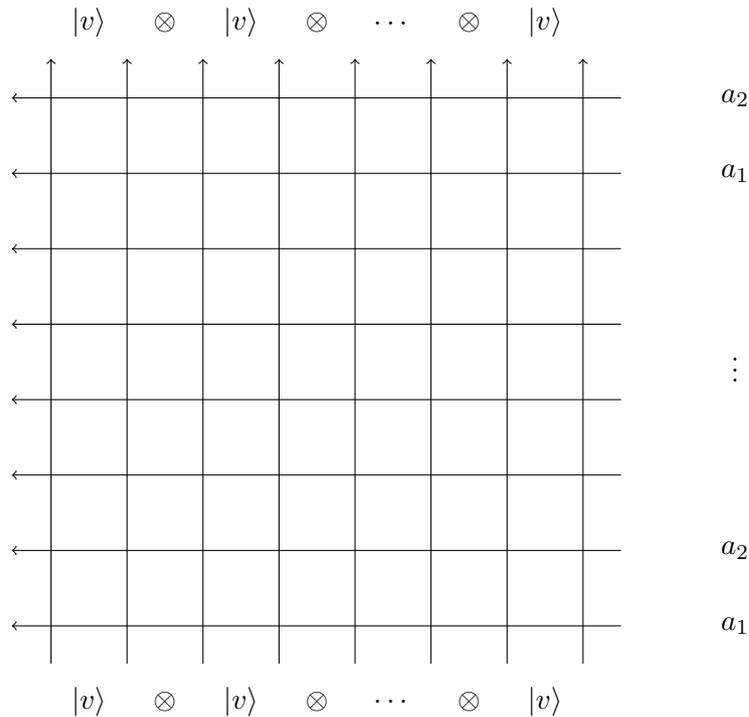

\centering
\begin{pgfpicture}{0cm}{0cm}{12cm}{9cm}

\pgfsetstartarrow{\pgfarrowto}

\pgfline{\pgfxy(1.5,1)}{\pgfxy(9.5,1)}
\pgfline{\pgfxy(1.5,2)}{\pgfxy(9.5,2)}
\pgfline{\pgfxy(1.5,3)}{\pgfxy(9.5,3)}
\pgfline{\pgfxy(1.5,4)}{\pgfxy(9.5,4)}
\pgfline{\pgfxy(1.5,5)}{\pgfxy(9.5,5)}
\pgfline{\pgfxy(1.5,6)}{\pgfxy(9.5,6)}
\pgfline{\pgfxy(1.5,7)}{\pgfxy(9.5,7)}
\pgfline{\pgfxy(1.5,8)}{\pgfxy(9.5,8)}

\pgfputat{\pgfxy(11,8)}{\pgfbox[center,center]{$a_2$}}
\pgfputat{\pgfxy(11,1)}{\pgfbox[center,center]{$a_1$}}
\pgfputat{\pgfxy(11,2)}{\pgfbox[center,center]{$a_2$}}
\pgfputat{\pgfxy(11,7)}{\pgfbox[center,center]{$a_1$}}
\pgfputat{\pgfxy(11,4.5)}{\pgfbox[center,center]{$\vdots$}}

\pgfline{\pgfxy(2,8.5)}{\pgfxy(2,0.5)}
\pgfline{\pgfxy(3,8.5)}{\pgfxy(3,0.5)}
\pgfline{\pgfxy(4,8.5)}{\pgfxy(4,0.5)}
\pgfline{\pgfxy(5,8.5)}{\pgfxy(5,0.5)}
\pgfline{\pgfxy(6,8.5)}{\pgfxy(6,0.5)}
\pgfline{\pgfxy(7,8.5)}{\pgfxy(7,0.5)}
\pgfline{\pgfxy(8,8.5)}{\pgfxy(8,0.5)}
\pgfline{\pgfxy(9,8.5)}{\pgfxy(9,0.5)}

\pgfputat{\pgfxy(2.5,9)}{\pgfbox[center,center]{$\ket{v}$}}
\pgfputat{\pgfxy(4.5,9)}{\pgfbox[center,center]{$\ket{v}$}}
\pgfputat{\pgfxy(6.5,9)}{\pgfbox[center,center]{$\dots$}}
\pgfputat{\pgfxy(8.5,9)}{\pgfbox[center,center]{$\ket{v}$}}
\pgfputat{\pgfxy(3.5,9)}{\pgfbox[center,center]{$\otimes$}}
\pgfputat{\pgfxy(5.5,9)}{\pgfbox[center,center]{$\otimes$}}
\pgfputat{\pgfxy(7.5,9)}{\pgfbox[center,center]{$\otimes$}}

\pgfputat{\pgfxy(2.5,0)}{\pgfbox[center,center]{$\ket{v}$}}
\pgfputat{\pgfxy(4.5,0)}{\pgfbox[center,center]{$\ket{v}$}}
\pgfputat{\pgfxy(6.5,0)}{\pgfbox[center,center]{$\dots$}}
\pgfputat{\pgfxy(8.5,0)}{\pgfbox[center,center]{$\ket{v}$}}
\pgfputat{\pgfxy(3.5,0)}{\pgfbox[center,center]{$\otimes$}}
\pgfputat{\pgfxy(5.5,0)}{\pgfbox[center,center]{$\otimes$}}
\pgfputat{\pgfxy(7.5,0)}{\pgfbox[center,center]{$\otimes$}}

\pgfclearendarrow

\end{pgfpicture}
\caption{The special partition function of the 6-vertex model which
  generates the Trotter decomposition of \eqref{GGs}.
There are $2N$ horizontal lines with rapidities equal to 0, 
and $L$ vertical lines with 
rapidities   $a_1=\beta/2N$ and $a_2=-\beta/2N+\eta$.
}
\label{naezaz}
\end{figure}

The transfer matrix is defined as the trace in auxiliary space
\begin{equation}
  t(u)=\text{Tr}_0\ T(u).
\end{equation}
It is known that $t(0)$ is proportional to the translation operator and the linear term in $u$ generates the Hamiltonian \cite{kluemper-review}. In the present normalization the following relation holds for large $N$:
\begin{equation*}
\frac{t(-\beta/2N)t(-\eta+\beta/2N)}{(\sinh(-\beta/2N+\eta))^{2L}}=1-\frac{\beta}{N}Q_2+\dots,
\end{equation*}
where
\begin{equation*}
  Q_2=\frac{1}{2\sinh(\eta)}H_{XXZ}.
\end{equation*}
Therefore
\begin{equation}
\label{hojj}
\bra{\Psi_0}\exp(-sH)\ket{\Psi_0}=
\lim_{N\to\infty} \frac{1}{(\sinh(-\beta/2N+\eta))^{2LN}}
\bra{\Psi_0}
\left(t(-\beta/2N)t(-\eta+\beta/2N)\right)^N
\ket{\Psi_0}  ,
\end{equation}
where
\begin{equation}
  \label{sbeta}
  \beta=2\sinh(\eta) s.
\end{equation}
The expression on the r.h.s. of 	\eqref{hojj} can be interpreted as the partition function of the 6-vertex
model with inhomogeneities and boundary conditions as given on figure
\ref{naezaz}. The transfer matrices in \eqref{hojj} correspond to
adding one more row the the diagram: they generate the imaginary time
evolution in the vertical direction.

\begin{figure}
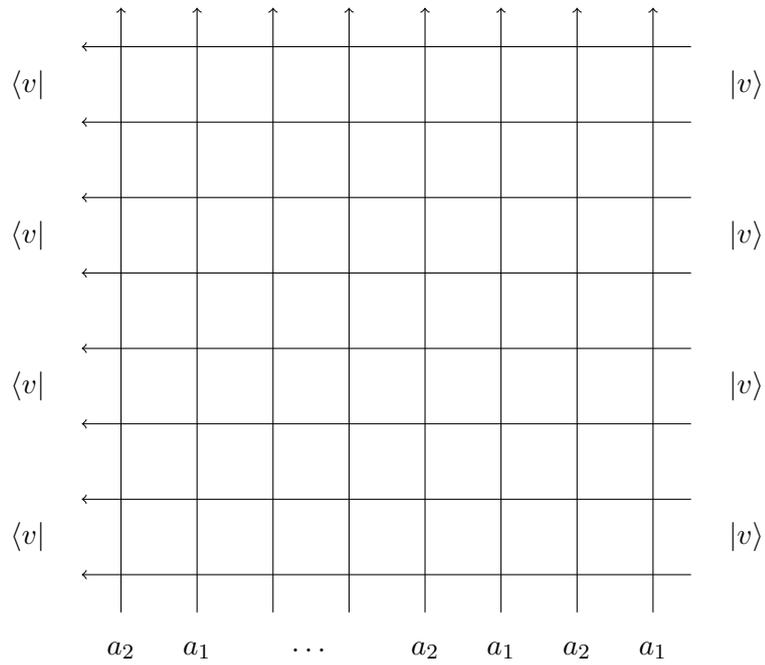

\centering
\begin{pgfpicture}{0cm}{4cm}{12cm}{13cm}

\pgfsetendarrow{\pgfarrowto}

\pgfline{\pgfxy(9.5,5)}{\pgfxy(1.5,5)}
\pgfline{\pgfxy(9.5,6)}{\pgfxy(1.5,6)}
\pgfline{\pgfxy(9.5,7)}{\pgfxy(1.5,7)}
\pgfline{\pgfxy(9.5,8)}{\pgfxy(1.5,8)}
\pgfline{\pgfxy(9.5,9)}{\pgfxy(1.5,9)}
\pgfline{\pgfxy(9.5,10)}{\pgfxy(1.5,10)}
\pgfline{\pgfxy(9.5,11)}{\pgfxy(1.5,11)}
\pgfline{\pgfxy(9.5,12)}{\pgfxy(1.5,12)}

\pgfclearstartarrow

\pgfline{\pgfxy(2,4.5)}{\pgfxy(2,12.5)}
\pgfline{\pgfxy(3,4.5)}{\pgfxy(3,12.5)}
\pgfline{\pgfxy(4,4.5)}{\pgfxy(4,12.5)}
\pgfline{\pgfxy(5,4.5)}{\pgfxy(5,12.5)}
\pgfline{\pgfxy(6,4.5)}{\pgfxy(6,12.5)}
\pgfline{\pgfxy(7,4.5)}{\pgfxy(7,12.5)}
\pgfline{\pgfxy(8,4.5)}{\pgfxy(8,12.5)}
\pgfline{\pgfxy(9,4.5)}{\pgfxy(9,12.5)}

\pgfputat{\pgfxy(9,4)}{\pgfbox[center,center]{$a_1$}}
\pgfputat{\pgfxy(8,4)}{\pgfbox[center,center]{$a_2$}}
\pgfputat{\pgfxy(7,4)}{\pgfbox[center,center]{$a_1$}}
\pgfputat{\pgfxy(6,4)}{\pgfbox[center,center]{$a_2$}}
\pgfputat{\pgfxy(4.5,4)}{\pgfbox[center,center]{$\dots$}}
\pgfputat{\pgfxy(3,4)}{\pgfbox[center,center]{$a_1$}}
\pgfputat{\pgfxy(2,4)}{\pgfbox[center,center]{$a_2$}}

\pgfclearendarrow

\pgfputat{\pgfxy(1,5.5)}{\pgfbox[right,center]{$ \bra{v}$}}
\pgfputat{\pgfxy(10,5.5)}{\pgfbox[left,center]{$ \ket{v}$}}
\pgfputat{\pgfxy(1,7.5)}{\pgfbox[right,center]{$ \bra{v}$}}
\pgfputat{\pgfxy(10,7.5)}{\pgfbox[left,center]{$ \ket{v}$}}
\pgfputat{\pgfxy(1,9.5)}{\pgfbox[right,center]{$ \bra{v}$}}
\pgfputat{\pgfxy(10,9.5)}{\pgfbox[left,center]{$ \ket{v}$}}
\pgfputat{\pgfxy(1,11.5)}{\pgfbox[right,center]{$ \bra{v}$}}
\pgfputat{\pgfxy(10,11.5)}{\pgfbox[left,center]{$ \ket{v}$}}

\end{pgfpicture}
\caption{The partition function after the reflection along the
  North-West diagonal.
 The rapidity associated to the
  horizontal lines is 0. The rapidities along the vertical lines are
  $a_1=-\beta/2N$ and $a_2=\beta/2N-\eta$.
Time evolution in the
vertical direction is generated by the Boundary Quantum Transfer
Matrix \eqref{mcT}, which adds two horizontal lines together with the boundary
conditions specified by $\bra{v}$ and $\ket{v}$ . }
\label{atirva1}
\end{figure}

This specific partition function can be evaluated alternatively by introducing
a new transfer matrix which acts in the other direction.
This is most easily achieved by performing a reflection along the
North-West diagonal (leaving the 6-vertex weights invariant) and leading
to the partition function depicted on figure \ref{atirva1}.
Here the horizontal lines define the action of 
 the so-called quantum transfer matrix:
\begin{equation*}
  T^{QTM}(u)=L_{2N,0}(u+\beta/2N)L_{2N-1,0}(u-\beta/2N+\eta)\dots 
L_{2,0}(u+\beta/2N)L_{1,0}(u-\beta/2N+\eta).
\end{equation*}
In the standard problem of the thermodynamics of the spin chain a similar partition function is obtained 
with periodic boundary conditions in both directions. Therefore in that case the
trace of $T^{QTM}$ in auxiliary space needs to be taken. However, for the quench problem considered here the boundary conditions on the left
and right side are non-trivial. This leads to
\begin{equation*}
  \bra{\Psi_0}\exp(-sH)\ket{\Psi_0}=
\lim_{N\to\infty}
 \text{Tr}\ \mathcal{T}^{L/2},
\end{equation*}
where
\begin{equation}
\label{mcT}
  \mathcal{T}=
 \frac{\bra{v}T^{QTM}(0)\otimes T^{QTM}(0)\ket{v} }
{(\sinh(-\beta/2N+\eta))^{4N}}.
\end{equation}
The scalar product in \eqref{mcT} is to be understood in the
tensor product of two auxiliary spaces. We call the operator $\mathcal{T}$ the Boundary Quantum Transfer Matrix.

Denoting the eigenvalues of $\mathcal{T}$ by $\Lambda_j$
\begin{equation}
\label{eee}
  \bra{\Psi_0}\exp(-sH)\ket{\Psi_0}=
\lim_{N\to\infty}
\sum_{j=1}^{2^{2N}}  (\Lambda_j)^{L/2}.
\end{equation}

In analogy with the periodic case \cite{kluemper-review} we make the following assumptions:
\begin{itemize}
\item There is a leading eigenvalue $\Lambda$ which remains separated
  from the others by a finite amount even in the $N\to\infty$ limit.
\item The large $L$ behaviour of \eqref{eee} can be studied by
  exchanging the limits $N\to\infty$ and $L\to\infty$.
\end{itemize}
The first assumption is justified by numerical checks (see below),
whereas we accept the second one based on experience with the periodic
case. Then the large volume behaviour is dominated by the leading eigenvalue:
\begin{equation*}
 \bra{\Psi_0}\exp(-sH)\ket{\Psi_0}\approx \left(\lim_{N\to\infty }\Lambda\right)^{L/2}  .
\end{equation*}
Finally
\begin{equation*}
  g(s)=\lim_{L\to\infty}\frac{1}{L}\log \bra{\Psi_0}\exp(-sH)\ket{\Psi_0}=
\frac{1}{2}\lim_{N\to\infty}\log \Lambda.
\end{equation*}
The remaining task is to diagonalize $\mathcal{T}$ and find its
leading eigenvalue in the Trotter limit. This can be achieved within the framework
of the boundary Algebraic Bethe Ansatz.

\section{Boundary Algebraic Bethe Ansatz}

The boundary Algebraic Bethe Ansatz was developed by Sklyanin
in \cite{sklyanin-boundary} to diagonalize Hamiltonians of open spin
chains with possible boundary magnetic fields. Here we use this
technology to diagonalize the Boundary Quantum Transfer Matrix. For a detailed
explanation of the method we refer the reader to \cite{openXXZ1}.

The boundary transfer matrix of a generic inhomogeneous spin chain of
length $M$ is defined as
\begin{equation}
\label{boundaryT}
  \mathcal{R}(u)=\text{Tr}_0\left\{ K^+(u) T_1(u)K^-(u)  T_2(u)\right\}.
\end{equation}
Here
\begin{equation*}
  T_1(u)=\tilde L_M(u)\dots \tilde L_1(u),
\end{equation*}
where $\tilde L_j(u)$ are local Lax-operators given by
\begin{equation*}
  \tilde L_j(u)=R_{0j}(u-\xi_j)
\end{equation*}
and 
\begin{equation}
  T_2(u)=\gamma(u) \sigma_0^y T^{t_0}_1(-u) \sigma_0^y,
\end{equation}
where $\gamma(u)=(-1)^M$. In the following $M$ is always even so we
are free to set $\gamma=1$. The parameters $\xi_j$ are the inhomogeneities along the chain. The boundary transfer matrix is depicted on figure \ref{startingpoint}.

For the $K$-matrices entering \eqref{boundaryT} we choose the diagonal
solution to the reflection equation \cite{sklyanin-boundary,openXXZ1}:
\begin{equation}
  K^\pm(u)=K(u\pm \eta/2,\xi_\pm)\quad\text{with}\quad
 K(u,\xi)=
 \begin{pmatrix}
   \sinh(\xi+u) & 0 \\
0 & \sinh(\xi-u)
 \end{pmatrix}
.
\end{equation}

Introducing the components of $T_1$ in auxiliary space as usual
\begin{equation}
  T_1(u)=
  \begin{pmatrix}
    A(u) & B(u) \\
C(u) & D(u)
  \end{pmatrix}
\end{equation}
the boundary transfer matrix can be written as
\begin{equation}
\label{Tukiirva}
\mathcal{R}(u)=  k_1^+k_1^- A(u)D(-u)-k_1^+k_2^- B(u)C(-u)-k^+_2k^-_1
  C(u)B(-u)+k^+_2k^-_2 D(u)A(-u),
\end{equation}
where
\begin{equation*}
  k_{a}^\pm \equiv K^{\pm}_{aa}
\end{equation*}
are the diagonal elements of the K-matrices.

The formula \eqref{Tukiirva} is equivalent to
\begin{equation}
\label{RR}
  \mathcal{R}(u)= 
 \bra{v^+(u)}T_1(u)\otimes T_1(-u)\ket{v^-(u)},
\end{equation}
where
\begin{equation}
  \ket{v^-(u)}=k_1^-(u)\ket{+-}-k_2^-(u) \ket{-+}\qquad
\bra{v^+(u)}=k_1^+(u)\bra{+-}-k_2^+(u) \bra{-+}.
\end{equation}
Setting  $u=0$ gives
\begin{equation}
\label{T00}
  \mathcal{R}(0)= 
 \bra{v^+}T_1(0)\otimes T_1(0)\ket{v^-}
\end{equation}
with 
\begin{equation}
\label{vpm}
\begin{split}
  \ket{v^-}&=\sinh(\xi^--\eta/2) \ket{+-}-
  \sinh(\xi^-+\eta/2)\ket {-+}\\
 \bra{v^+}&=\sinh(\xi^++\eta/2) \bra{+-}-
  \sinh(\xi^+-\eta/2)\bra {-+}.
\end{split}
\end{equation}
The boundary transfer matrix \eqref{T00} is proportional to the
operator $\mathcal{T}$ of \eqref{mcT} if the following identifications
are made:
\begin{itemize}
\item $M=2N$.
\item The inhomogeneities are $\xi_{2j+1}=-\beta/2N$,
  $\xi_{2j}=\beta/2N-\eta$.
\item The parameters $\xi^\pm$ of the $K$-matrices are determined by
  \begin{equation}
    \gamma=\frac{\sinh(\xi^-+\eta/2)}{\sinh(\xi^--\eta/2)}=
\frac{\sinh(\xi^+-\eta/2)}{\sinh(\xi^++\eta/2)}
  \end{equation}
\end{itemize}
If the above conditions hold then
\begin{equation}
\label{TR}
\begin{split}
&  \mathcal{T}=\frac{1}{(\sinh(-\beta/2N+\eta))^{4N}}
\frac{1}{\skalarszorzat{v^+}{v^-}}
\mathcal{R}(0).
\end{split}
\end{equation}

The common eigenstates of the operators $\mathcal{R}$ can be created
from the ferromagnetic reference state $\ket{F}=\ket{++\dots}$ as
\begin{equation}
\label{bABAstates}
  \ket{\{\lambda\}_n}=\prod_{j=1}^n \mathcal{B}_-(\lambda_j) \ket{F},
\end{equation}
where the $\mathcal{B}_-(\lambda)$ operators are defined through
\begin{equation*}
  \mathcal{U}_-(\lambda)=
T_1(\lambda) K^-(\lambda) T_2(\lambda)=
  \begin{pmatrix}
   \mathcal{A}_-(u) & \mathcal{B}_-(u) \\
\mathcal{C}_-(u) & \mathcal{D}_-(u)
  \end{pmatrix}.
\end{equation*}
The states in \eqref{bABAstates} are eigenstates if the rapidities
satisfy the Bethe equations
\begin{equation}
  \label{BBA}
\begin{split}
&
\left[
\frac{\sinh(\lambda_j+\beta/2N-\eta)}{\sinh(\lambda_j-\beta/2N+\eta)}
\frac{\sinh(\lambda_j-\beta/2N)}{\sinh(\lambda_j+\beta/2N)}
\right]^{2N}
\prod_{k\ne j}
\frac{\sinh(\lambda_j-\lambda_k+\eta)\sinh(\lambda_j+\lambda_k+\eta)}
{\sinh(\lambda_j-\lambda_k-\eta)\sinh(\lambda_j+\lambda_k-\eta)}\times
\\
&\times\frac{\sinh(\lambda_j-(\xi_+-\eta/2))}{\sinh(\lambda_j+(\xi_+-\eta/2))}
\frac{\sinh(\lambda_j-(\xi_--\eta/2))}{\sinh(\lambda_j+(\xi_--\eta/2))}=1.
\end{split}
\end{equation}
For every set of Bethe roots $\{\lambda\}_n$ it is useful to introduce
the doubled set
$\{\tilde \lambda\}_{2n}=\{\lambda\}_n\cup \{-\lambda\}_n$. Then the
Bethe equations read
\begin{equation}
  \label{BBA2}
\begin{split}
&
\left[
\frac{\sinh(\lambda_j+\beta/2N-\eta)}{\sinh(\lambda_j-\beta/2N+\eta)}
\frac{\sinh(\lambda_j-\beta/2N)}{\sinh(\lambda_j+\beta/2N)}
\right]^{2N}
\prod_{k=1}^{2n}
\frac{\sinh(\lambda_j-\tilde \lambda_k+\eta)}{\sinh(\lambda_j-\tilde \lambda_k-\eta)}
\times
\\
&\times\frac{\sinh(\lambda_j-(\xi_+-\eta/2))}{\sinh(\lambda_j+(\xi_+-\eta/2))}
\frac{\sinh(\lambda_j-(\xi_--\eta/2))}{\sinh(\lambda_j+(\xi_--\eta/2))}
\frac{\sinh(2\lambda_j-\eta)}{\sinh(2\lambda_j+\eta)}  =-1.
\end{split}
\end{equation}
The eigenvalues $\Lambda(u,\{\lambda\}_n) $of the transfer matrix
$\mathcal{R}(u)$ are given by
\begin{equation}
\begin{split}
\Lambda(u)=&
\frac{1}{\sinh(2u)}\Big[\sinh(2u+\eta)\sinh(u+\xi^--\eta/2)\sinh(u+\xi^--\eta/2)\times\\
&\hspace{2cm}\times (\sinh(u-\beta/2N+\eta)\sinh(u+\beta/2N))^{2N}
\prod_{k=1}^{2n}
\frac{\sinh(u-\tilde\lambda_k-\eta)}{\sinh(u-\tilde\lambda_k)}
\\
&
+\sinh(2u-\eta)\sinh(u+\xi^-+\eta/2)\sinh(u+\xi^-+\eta/2)\times\\
&\hspace{2cm}\times (\sinh(u+\beta/2N-\eta)\sinh(u-\beta/2N))^{2N}
\prod_{k=1}^{2n} \frac{\sinh(u-\tilde\lambda_k+\eta)}{\sinh(u-\tilde\lambda_k)}
\Big].
\end{split}
\end{equation}
These equations are the basis for analyzing the leading eigenvalue $\Lambda$ of the operator $\mathcal{T}$. In the following section we focus on the particular case of the N\'eel state for generic $\Delta$. In section \ref{sec:dimer} the case of the dimer state is considered for $\Delta=1$.

\section{Boundary QTM: The N\'eel state}

Putting 
\begin{equation*}
  \xi^-=-\eta/2,\quad
 \xi^+=\eta/2
\end{equation*}
into \eqref{vpm} leads to 
\begin{equation}
\ket{v^-}=-\sinh(\eta)\ket{+-}\qquad
\bra{v^+}=\sinh(\eta)\bra{+-}.
\end{equation}
These two-site states generate the N\'eel state with the normalization following from
\begin{equation}
\skalarszorzat{v^+}{v^-}=-\sinh^2(\eta).
\end{equation}
From \eqref{BBA2} follow the Bethe equations
\begin{equation}
  \label{BBA3}
\begin{split}
K(\lambda_j)
\left[
\frac{\sinh(\lambda_j+\beta/2N-\eta)}{\sinh(\lambda_j-\beta/2N+\eta)}
\frac{\sinh(\lambda_j-\beta/2N)}{\sinh(\lambda_j+\beta/2N)}
\right]^{2N}
\prod_{k=1}^{2n}
\frac{\sinh(\lambda_j-\tilde\lambda_k+\eta)}{\sinh(\lambda_j-\tilde\lambda_k-\eta)}
=-1
\end{split}
\end{equation}
with
\begin{equation*}
  K(u)=\frac{\sinh(u+\eta)}{\sinh(u-\eta)}
\frac{\sinh(2u-\eta)}{\sinh(2u+\eta)}.
\end{equation*}
The eigenvalues of $\mathcal{R}(u)$ at $u=0$ take the remarkably
simple form
\begin{equation}
\label{L0}
\begin{split}
\Lambda(0)=
-\sinh^2(\eta)
(\sinh(-\beta/2N+\eta)\sinh(\beta/2N))^{2N}
\times
\prod_{k=1}^{2n}
\frac{\sinh(\tilde\lambda_k+\eta)}{\sinh(\tilde\lambda_k)}.
\end{split}
\end{equation}

These equations will be analyzed further in the case of $\Delta>1$ corresponding to $\eta\in\valos$.
Experience with the periodic case suggests that at any finite $N$ the leading eigenvalue
will be given by a state with $N$ Bethe roots which are all situated
at the imaginary axis. We checked that this is
indeed true: We constructed the matrix $\mathcal{R}(0)$ using the
computer program \texttt{octave} and diagonalized it numerically for
small systems of $N=2,4,6,8$. We searched for the unique solution of
\eqref{BBA3} with $N$ roots at the imaginary axis. Computing \eqref{L0} we found exact
agreement with the result of exact diagonalization. Moreover we found
that the behaviour of the next to leading eigenvalue is consistent
with a non-vanishing gap in the $N\to\infty$ limit.

Collecting the normalization factors 
 the leading eigenvalue of $\mathcal{T}$ is 
\begin{equation}
\begin{split}
\Lambda=
\left(\frac{\sinh(\beta/2N)}{\sinh(\eta-\beta/2N)}\right)^{2N}
\times
\prod_{k=1}^{2N}
\frac{\sinh(\tilde\lambda_k+\eta)}{\sinh(\tilde\lambda_k)}.
\end{split}
\end{equation}

Solving the Bethe equations showed that the
behaviour of the roots as a function of $N$ is the same as in the
periodic case: they cluster around $u=0$ but they do not become
dense at any $u\ne 0$, all roots have a finite limit as $N\to\infty$.

\subsection{Taking the Trotter limit}

We define the auxiliary function
\begin{equation}
\begin{split}
  \fa(u)=
K(u)
\left[
\frac{\sinh(u+\beta/2N-\eta)}{\sinh(u-\beta/2N+\eta)}
\frac{\sinh(u-\beta/2N)}{\sinh(u+\beta/2N)}
\right]^{2N}
\prod_{k=1}^{2N} \frac{\sinh(u-\tilde \lambda_k+\eta)}{\sinh(u-\tilde \lambda_k-\eta)}.
\end{split}
\end{equation}
Also, we define
\begin{equation}
\label{Adef}
  A(u)=\frac{1+\fa(u)}{1+K(u)}.
\end{equation}
Both functions are $i\pi$ periodic. 

We define the canonical contour $C$ just as in the periodic case: it has to encircle all Bethe roots but no additional zeroes of $1+\fa(u)$ \cite{kluemper-review}. For $\Delta>1$ the contour can be chosen to consist of two vertical line segments running from $\alpha-i\pi/2$ to $\alpha+i\pi/2$ and from $-\alpha+i\pi/2$ to $-\alpha-i\pi/2$. Here $\alpha\in\valos$ is an arbitrary parameter satisfying $\alpha<\eta/2$.

The analytic properties of $A(u)$ inside the contour $C$ are as follows:
\begin{itemize}
\item It has an $2N$th order pole at $u=\beta/2N$.
\item It has $2N$ zeroes at the (doubled set of) Bethe roots
  $\tilde\lambda_j$.
\item It has no additional zeroes or poles, therefore its winding
  number is zero and its logarithm can be defined to be single valued.
\end{itemize}

Note that $1+\fa(u)$ has an extra zero at $u=i\pi/2$, however this is
canceled by the denominator of \eqref{Adef}. We checked with
\texttt{Mathematica} that for the solutions of the Bethe equations
with $n=N$ there are indeed no other zeroes within the canonical contour.

For any function $f(\omega)$ which is analytic 
within the contour the following holds
\begin{equation*}
\int_C\frac{d\omega}{2\pi i}f'(\omega) \log(A(\omega))=
2Nf(-\beta/2N)-\sum_{j=1}^{2N} f(\tilde\lambda_j).
\end{equation*}
Taking 
\begin{equation*}
  f(\omega)=\log\frac{\sinh(u-\omega+\eta)}{\sinh(u-\omega-\eta)}
\end{equation*}
leads to the non-linear integral equation (NLIE)
\begin{equation*}
\begin{split}
  \log \fa(u)=
&\log(K(u)) +
2N \log\left(
\frac{\sinh(u-\beta/2N)}{\sinh(u+\beta/2N)}
\frac{\sinh(u+\beta/2N+\eta)}{\sinh(u-\beta/2N+\eta)}
\right)-\\
& 
-\int_C\frac{d\omega}{2\pi i}
\frac{\sinh(2\eta)}{\sinh(u-\omega+\eta)\sinh(u-\omega-\eta)}
 \log(A(\omega)).
\end{split}
\end{equation*}
Similarly
\begin{equation*}
  \log \Lambda=
\int_C\frac{d\omega}{2\pi i}
\frac{\sinh\eta}{\sinh(\omega)\sinh(\omega+\eta)}
 \log(A(\omega)).
\end{equation*}
Here we also used the fact that $\log(A(0))=0$.

The above equations are valid at any $N$ and they have a well-behaving
Trotter limit. For the auxiliary function we obtain
\begin{equation}
\label{BQTMhoho}
\begin{split}
  \log \fa(u)=
\log(K(u))
-4s \frac{\sinh^2\eta}{\sinh(u)\sinh(u+\eta)}
-\int_C\frac{d\omega}{2\pi i}
\frac{\sinh(2\eta)}{\sinh(u-\omega+\eta)\sinh(u-\omega-\eta)}
 \log(A(\omega)).
\end{split}
\end{equation}
where we used the relation \eqref{sbeta}.

Finally for the dynamical free energy density we obtain
\begin{equation}
\label{BQTMhehe}
  g(s)=\frac{1}{2}\log \Lambda=
\frac{1}{2}\int_C\frac{d\omega}{2\pi i}
\frac{\sinh\eta}{\sinh(\omega)\sinh(\omega+\eta)}
 \log(A(\omega)).
\end{equation}
Equations \eqref{BQTMhoho}-\eqref{BQTMhehe} constitute the main
result of this section. They are valid for any $s\in\valos^+$ and
they can be used as a basis for the analytic continuation $s=it$ which
is investigated in subsection \ref{sec:Loschmidt}.

\subsection{The small $s$ limit}

It is useful to check the $s\to 0$ limit of the NLIE analytically, as
this provides a non-trivial check of the calculations. The solution of the NLIE at $s=0$ is
\begin{equation*}
  \fa(u)=K(u)\qquad A(u)=1,
\end{equation*}
which gives $g(0)=0$ as it should by its definition.

The first order term in $g(s)$ is expected to be the Hamiltonian density:
\begin{equation}
  g(s)=-\frac{\bra{N}H\ket{N}}{L}=2\cosh\eta.
\end{equation}
In the following we derive this result from the NLIE. We define
\begin{equation*}
  \fa'(u)=\frac{1}{\fa(u)}\frac{\partial \fa(u)}{\partial s}.
\end{equation*}
This function satisfies the linear integral equation
\begin{equation*}
\begin{split}
  \fa'(u)=
-4 \frac{\sinh^2\eta}{\sinh(u)\sinh(u+\eta)}
-\int_C\frac{d\omega}{2\pi i}
\frac{\sinh(2\eta)}{\sinh(u-\omega+\eta)\sinh(u-\omega-\eta)}
\frac{\fa'(u)\fa(u)}{1+\fa(u)}.
\end{split}
\end{equation*}
The solution at $s=0$ can be obtained by simple contour integrals, leading to
\begin{equation*}
\fa'(\lambda)= 
4\frac{\sinh^2(\eta)\cosh(\lambda)}{\sinh(\lambda)\sinh(\lambda-\eta)\sinh(\lambda+\eta)}  .
\end{equation*}

Finally we obtain
\begin{equation*}
\begin{split}
\left.\frac{\partial g}{\partial s}\right|_{s=0}&=
\int_C\frac{d\omega}{2\pi i}
\frac{\sinh^2\eta}{\sinh(\omega)\sinh(\omega+\eta)}
 \frac{\fa'(\omega)}{1+K(-u)}\\
&=\int_C\frac{d\omega}{2\pi i}
\frac{2\sinh^4\eta \cosh(\omega)}
{\sinh^2(\omega)\sinh(\omega+\eta)\sinh(\omega-\eta)}\times
\\
&\hspace{3cm}
\times\frac{\sinh(2u-\eta)}{\sinh(u+\eta)\sinh(2u-\eta)+\sinh(u-\eta)\sinh(2u+\eta)}\\
&=2\cosh\eta.
\end{split}
\end{equation*}

It is also possible to derive the higher order terms from the NLIE by
taking further derivatives and solving linear equations. However, this
becomes very cumbersome already for the second cumulant, so in
practise it is more convenient to determine them from explicit real
space calculations. For the sake of completeness we give here the
second cumulant:
\begin{equation}
\label{kappa2}
  \kappa_2=\frac{\bra{N}H^2\ket{N}-\bra{N}H\ket{N}^2}{L}=4.
\end{equation}

\subsection{The large $s$ limit}

At large positive $s$ the behaviour of $g(s)$ will be determined by
the low lying states of the antiferromagnetic Hamiltonian. In the $\Delta>1$ regime
considered here the 
two lowest lying states of the Hamiltonian $\ket{GS_1}$ and $\ket{GS_2}$ are such that they become
degenerate in the thermodynamic limit with an energy density $e_0$, but a finite gap remains between
them and the next state. Therefore in the large $s$ limit we have
\begin{equation}
  e^{g(s)L}\approx \left(|\skalarszorzat{N}{GS_1}|^2+|\skalarszorzat{N}{GS_2}|^2\right)e^{-se_0L}.
\end{equation}
The overlap of the ground states with the N\'eel state scales as
\begin{equation}
  |\skalarszorzat{N}{GS_{1,2}}|^2=\alpha_{1,2} \exp(\beta_{1,2}L).
\end{equation}
Although the pre-factors $\alpha_1$ and $\alpha_2$ can be different,
we expect that the exponent is the same:
$\beta_1=\beta_2=\beta$. We checked this by a finite volume numerical investigation, which will be published elsewhere. We call the quantity $e^\beta$ the
``overlap per site''.

Putting everything together the large $s$ behaviour of $g(s)$ is
\begin{equation}
\label{asy}
  g(s)=-e_0 s+\beta+\dots
\end{equation}
In the following we extract $e_0$ and $\beta$ from the NLIE. First we
perform a rotation of $\pi/2$ in the complex plain and introduce
\begin{equation}
  \tilde\fa(u)=\frac{\fa(u)}{K(u)}.
\end{equation}
This way the NLIE takes the form
\begin{equation}
\label{BNLIE-rot}
\begin{split}
\log \tilde\fa(\lambda)=&
s \frac{4\sinh^2\eta}{\sin(\lambda)\sin(\lambda-i\eta)}
+  \int_C \frac{d\omega}{2\pi } 
\frac{2\sinh2\eta}{\sin(\lambda-\omega+i\eta)\sin(\lambda-\omega-i\eta)}  \log(A(\omega))
\end{split}
\end{equation}
with
\begin{equation}
  A(u)=\frac{1+K(u)\tilde\fa(u)}{1+K(u)}
\end{equation}
The integration contour consists of two horizontal line segments
\begin{equation}
 C^+=[-\pi/2+i\alpha\dots \pi/2+i\alpha] \quad\text{and}\quad
 C^-= [-\pi/2-i\alpha\dots \pi/2-i\alpha]
\end{equation}
For the dynamical free energy we obtain
\begin{equation}
  \label{g-qtm-rot}
  g(s)=-\frac{1}{2}\int_{C}\frac{d\omega}{2\pi } 
\frac{\sinh\eta \log(A(\omega))}{\sin(\lambda)\sin(\lambda-i\eta)}.
\end{equation}
Note that as a complex integral $d\omega$ is negative on the upper
and positive on the lower contour, ie. formally we have
\begin{equation*}
  \int_{C}\frac{d\omega}{2\pi } =
-\int_{C^+}\frac{d\omega}{2\pi } +
\int_{C^-}\frac{d\omega}{2\pi } 
\end{equation*}
where the integrals on the r.h.s. are to be understood as purely real integrals.

The auxiliary function can be expanded as
\begin{equation*}
  \log \tilde \fa(\lambda)=s \rho(\lambda)+\kappa(\lambda)+\dots
\end{equation*}
Investigating the numerical solutions of the NLIE we find that
the real part of $\rho$ is positive on the upper contour, and negative
on the lower contour. Therefore on the upper contour
\begin{equation}
  \log A(u)\quad \to \quad s \rho(\lambda)+\kappa(\lambda)+\log\frac{K(u)}{1+K(u)}
\end{equation}
whereas on the lower contour
\begin{equation}
  \log A(u)\quad \to \quad \log\frac{1}{1+K(u)}
\end{equation}

First we compute the linear terms in $s$ for which only the upper contour contributes. For
$\rho(u)$ we obtain the integral equation
  \begin{equation}
\label{BNLIE-rho}
\begin{split}
 \rho(\lambda)=&
 \frac{4\sinh^2\eta}{\sin(\lambda)\sin(\lambda-i\eta)}
-  \int_{C^+} \frac{d\omega}{2\pi } 
\frac{\sinh 2  \eta}{\sin(\lambda-\omega+i\eta)\sin(\lambda-\omega-i\eta)}  \rho(\omega).
\end{split}
\end{equation}
This integral can be solved in Fourier space and leads to
\begin{equation*}
  e_0=\frac{1}{2}\int_{C^+}\frac{d\omega}{2\pi } 
\frac{\sinh\eta}{\sin(\lambda)\sin(\lambda-i\eta)}  \rho(\lambda)=
2\sinh\eta\sum_{n=-\infty}^\infty \frac{e^{-\eta |n|}}{\cosh(\eta n)}.
\end{equation*}
This is the known formula for the ground state energy of the XXZ chain
in the massive regime.

For the sub-leading contributions we need to keep the $\ordo(1)$ terms
from the lower contour too. For $\kappa(u)$ we obtain the linear equation
\begin{equation}
\label{BNLIE-kappa}
\begin{split}
 \kappa(\lambda)=&
-  \int_{C^+} \frac{d\omega}{2\pi } 
\frac{\sinh 2\eta}{\sin(\lambda-\omega+i\eta)\sin(\lambda-\omega-i\eta)}  
\kappa(\omega)\\
&-  \int_{C^+} \frac{d\omega}{2\pi } 
\frac{\sinh 2\eta}{\sin(\lambda-\omega+i\eta)\sin(\lambda-\omega-i\eta)}  
\log\frac{K(i\omega)}{1+K(i\omega)}\\
&-\int_{C^-} \frac{d\omega}{2\pi } 
\frac{\sinh 2\eta}{\sin(\lambda-\omega+i\eta)\sin(\lambda-\omega-i\eta)}  
\log(1+K(i\omega)).
\end{split}
\end{equation}
Finally for the exponent $\beta$ we get
\begin{equation}
\label{ghj}
\begin{split}
  \beta=
\frac{1}{2}\int_{C^+}\frac{d\omega}{2\pi } 
\frac{\sinh\eta}{\sin(\lambda)\sin(\lambda-i\eta)}\kappa(\omega)
&+\frac{1}{2}\int_{C^+}\frac{d\omega}{2\pi } 
\frac{\sinh\eta
}{\sin(\lambda)\sin(\lambda-i\eta)}\log\frac{K(i\omega)}{1+K(i\omega)}\\
&-\frac{1}{2}\int_{C^-}\frac{d\omega}{2\pi } 
\frac{\sinh\eta
}{\sin(\lambda)\sin(\lambda-i\eta)}\log\frac{1}{1+K(i\omega)}.
\end{split}
\end{equation}
Equations \eqref{BNLIE-kappa}-\eqref{ghj} can be solved easily in
Fourier-space,  numerical results 
are shown in Table \ref{tab:ops}. Generally we observe that the
overlap per site is a monotonically growing function of $\Delta$, in
the $\Delta=\infty$ limit it approaches 1 as expected, and it has a
finite limit at $\Delta=1$.

\begin{table}
\centering
\begin{tabular}{|c||c|c|c|c|c|}
\hline
$\Delta$  & 1.01 & 1.1 & 1.5 & 2 &  4 \\
\hline
$\exp(\beta)$ & 0.83602143 & 0.84896360 & 0.90296103  & 0.94168383  &  0.98462308  \\
\hline
\end{tabular}
\caption{The overlap per site between the N\'eel state and the ground state of the XXZ Hamiltonian for different $\Delta>1$.}
\label{tab:ops}
\end{table}

\subsection{Numerical results for $g(s)$ and the Loschmidt echo}

\label{sec:Loschmidt}

\begin{figure}
 \centering
\subfigure[The dynamical free energy density for real $s$. The solid line shows the results of the NLIE, whereas the straight line is the large $s$ asymptotic \eqref{asy}. ]{\includegraphics{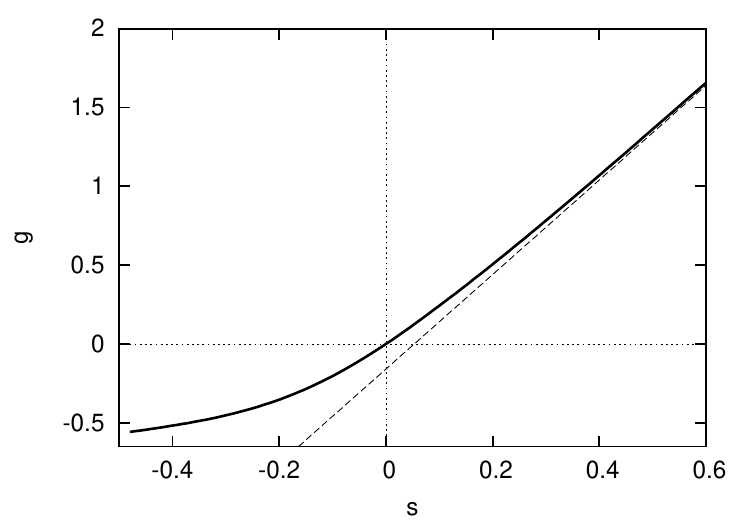}  }
\hspace{0.2cm}
\subfigure[The Loschmidt-echo per site for small real times. The squares represent the results of the NLIE, whereas the dashed line shows the quadratic approximation.]{\includegraphics{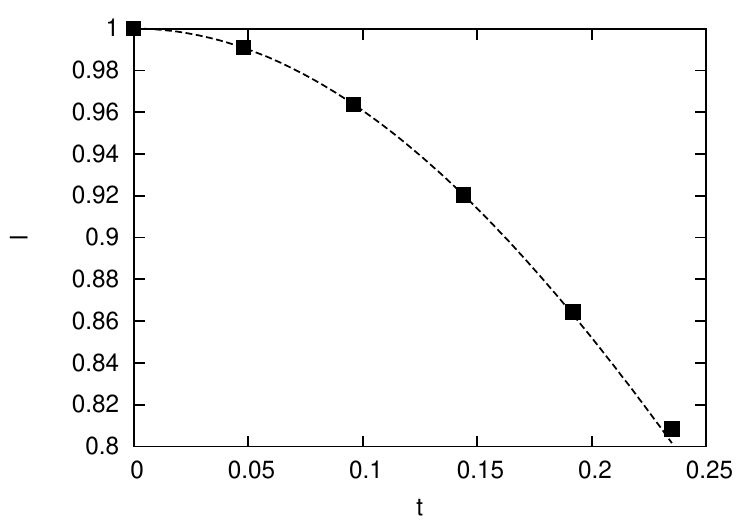}  }
 \caption{Numerical results for the dynamical free energy density and the Loschmidt echo for $\eta=0.5$ ($\Delta=1.1276$).}
 \label{fig:gs}
\end{figure}

We numerically implemented the NLIE \eqref{BQTMhoho}-\eqref{BQTMhehe}
for $\Delta>1$ using the computer program 
\texttt{octave}. As a first step we investigated the behaviour of
$g(s)$ for real positive $s$. It was found that the simple iteration
technique converges and the results do not depend on the parameter
$\alpha$ determining the position of the integration contour. Numerical results
(together with the predicted large $s$ asymptotic) are shown in Figure
\ref{fig:gs} (a). We also investigated the position of the zeroes of
the function $A(u)$ defined by
\eqref{Adef}. It was found that for purely real values of  $s$ all
zeroes lie symmetrically on the imaginary axis, as expected.

As a second step we turned to the problem of analytic continuation. It
was found that the NLIE remains stable in a finite neighbourhood of
the positive real axis. Giving small imaginary parts to $s$ we
observed that the zeroes of $A(u)$ move away from the imaginary
axis, but they still cluster at $u=0$. The NLIE remains stable even
for $s=it$, $t\in\valos$ up to a certain critical value $t=t^*$. In this case all
zeroes of $A(u)$ are positioned on the real axis. We observed that the
zeroes move outwards as $t$ is increased and they approach the contour
$C$. The NLIE is valid until all zeroes lie within the
contour\footnote{This statement can be proven by the following argument.
The zeroes of $A(u)$ cluster around $u=0$, but fixing a specific zero
it can be shown that its position depends continuously on the
  Trotter number $N$. Therefore, if a zero is outside of the
canonical contour in the Trotter limit, then this means that there is
an $N_c$ such that for all $N>N_c$ it is outside the contour and then
the NLIE is not valid anymore.
}, therefore it is useful to choose the maximal value
$\alpha=\eta/2-\eps$.
However, at the threshold $t=t^*$ the NLIE becomes numerically
unstable even before the outermost zero crosses the contour.
We found that
the critical value is approximately $t^*\approx \eta/(2\sinh\eta)$. In Fig. \ref{fig:gs} (b) we
plot the numerical results for the Loschmidt echo per site for
$\eta=0.5$ and $0<t<t^*$. In this regime the quadratic approximation
using the second cumulant \eqref{kappa2}
works very well and it deviates only slightly from the exact result of the NLIE.

Based on the behaviour of the zeroes 
in the regime  $0<t<t^*$
 we expect that for all $s=it$ 
they move
outwards as $t$ is increased, and they leave the canonical contour one
by one at
certain threshold values $t^*_j$, $j=1\dots\infty$. The structure of
the NLIE has to be changed at each of these points to account for the
missing zeroes. 

In the numerical investigations above we assumed that the gap of the
BQTM does not vanish and that the leading eigenvalue is always given
by the analytic continuation of the same state. This certainly holds if $s$ is close enough to the
real positive axis. However, for general complex $s$ level crossings
can appear which lead to non-analytic behaviour in the dynamical free energy.
We plan to address these questions in a future work.

\section{Boundary QTM: The XXX limit and the dimer state}

\label{sec:dimer}

In the present section we consider the XXX chain where $\Delta=1$. 
 Typically all formulas relevant for the Bethe Ansatz
solution of the XXX chain are easily obtained from their XXZ
counterparts by a simple $\eta\to 0$ limit and a rescaling of the
rapidities by $\eta$ \footnote{This is true for formal expressions
  about individual states or the thermodynamic quantities. The behaviour
of correlation functions is of course different, because the XXX
chain is critical, whereas the XXZ chain with $\Delta>1$ is
massive.}. This is also true for the quantities considered in the
present work. Here we only give the main equations and do
not repeat the whole derivation presented in the previous two
sections. We will be mainly concerned with the dimer state, ie. we
calculate  $g(s)$ defined as
\begin{equation*}
  e^{g(s)L}=\bra{D} e^{-sH}\ket{D},
\end{equation*}
where
\begin{equation*}
  D=\otimes_{j=1}^{L/2} \left(\frac{\ket{+-}-\ket{-+}}{\sqrt{2}}\right).
\end{equation*}

The algebraic Bethe Ansatz construction uses the rational R-matrix
\begin{equation}
  R(u)=
  \begin{pmatrix}
    u+i & & &\\
& u  & i & \\
& i & u & \\
& & & u+i
  \end{pmatrix}.
\label{Rxxx}
\end{equation}
The boundary $K$-matrices with generic parameters $\xi_\pm$ are
\begin{equation}
\label{XXXK}
K_\pm(u)=K(u\pm i/2,\xi_\pm)\quad\text{with}\quad
 K(u,\xi)=
 \begin{pmatrix}
   \xi+u & 0 \\
0 & \xi-u
 \end{pmatrix}.
\end{equation}
The main objective is to diagonalize the matrix
\begin{equation}
\label{TRXXX}
\begin{split}
&  \mathcal{T}=\frac{1}{(-i\beta/2N+i)^{4N}}
\frac{1}{\skalarszorzat{v^+}{v^-}}
\mathcal{R}(0),
\end{split}
\end{equation}
where $\mathcal{R}(u)$ is the boundary transfer matrix
\begin{equation}
\label{RRXXX}
  \mathcal{R}(u)= 
 \bra{v^+(u)}T_1(u)\otimes T_1(-u)\ket{v^-(u)}.
\end{equation}
The $K$-matrices \eqref{XXXK} lead to the following two-site boundary states:
\begin{equation}
\begin{split}
  \ket{v^-}&\equiv \ket{v^-(0)}=(\xi^--i/2) \ket{+-}-
  (\xi^-+i/2)\ket {-+}\\
 \bra{v^+}&\equiv \bra{v^+(0)}=(\xi^++i/2) \bra{+-}-
  (\xi^+-i/2)\bra {-+}.
\end{split}
\end{equation}
The Bethe Ansatz equations for a set of roots $\{\lambda\}_n$ are
\begin{equation}
  \label{BBA2XXX}
\begin{split}
&
\left[
\frac{(\lambda_j+i \beta/2N-i)}{(\lambda_j-i \beta/2N+i)}
\frac{(\lambda_j-i \beta/2N)}{(\lambda_j+i \beta/2N)}
\right]^{2N}
\prod_{k=1}^{2n}
\frac{(\lambda_j-\tilde\lambda_k+i)}{(\lambda_j-\tilde\lambda_k-i)}\times
\\
&\times
\frac{(\lambda_j-(\xi_+-i/2))}{(\lambda_j+(\xi_+-i/2))}
\frac{(\lambda_j-(\xi_--i/2))}{(\lambda_j+(\xi_--i/2))}
\frac{(2\lambda_j-i)}{(2\lambda_j+i)}
=-1
\end{split}
\end{equation}
and the eigenvalue of $\mathcal{R}(u)$ on the given state is
\begin{equation}
\begin{split}
\Lambda(u)=&
\frac{1}{2u}\left[(2u+i)(u+\xi_+-i/2)(u+\xi_--i/2)
((u-i \beta/2N+i)(u+i \beta/2N))^{2N}
\times
\prod_{k=1}^{2n}
\frac{(u-\tilde\lambda_k-i)}{(u-\tilde\lambda_k)}\right.
\\
&\left.
+(2u-i)(u-\xi_++i/2)(u-\xi_-+i/2)
((u+i \beta/2N-i)(u-i \beta/2N))^{2N}
\times
\prod_{k=1}^{2n} \frac{(u-\tilde\lambda_k+i)}{(u-\tilde\lambda_k)}
\right].
\end{split}
\end{equation}

The case of the N\'eel state is obtained by setting $\xi_-=-i/2$ and
$\xi^+=i/2$. The resulting equations follow from a  straightforward limit of
the XXZ formulas, therefore we do not consider this case in detail. 
Instead we focus on the limit
\begin{equation*}
  \xi_\pm \to \infty
\end{equation*}
which produces the dimer states as
\begin{equation}
\begin{split}
  \lim_{\xi^-\to\infty }\frac{\ket{v^-}}{\sqrt{2} \xi^-}
=\lim_{\xi^+\to\infty }\frac{\ket{v^+}}{\sqrt{2} \xi^+}=\frac{1}{\sqrt{2}}(\ket{+-}-\ket {-+}).\\
 \end{split}
\end{equation}
In this limit the Bethe equations become
\begin{equation}
  \label{BBA2XXXdimer}
\begin{split}
K_D(\lambda_j)
\left[
\frac{(\lambda_j+i \beta/2N-i)}{(\lambda_j-i \beta/2N+i)}
\frac{(\lambda_j-i \beta/2N)}{(\lambda_j+i \beta/2N)}
\right]^{2N}
\prod_{k=1}^{2N}
\frac{(\lambda_j-\tilde\lambda_k+i)}{(\lambda_j-\tilde\lambda_k-i)}
=-1
\end{split}
\end{equation}
with
\begin{equation*}
  K_D(u)=\frac{(2u-i)}{(2u+i)}.
\end{equation*}
The eigenvalues of $\mathcal{T}$ are 
\begin{equation}
\label{LaD}
\begin{split}
\Lambda=&
\frac{1}{(-i\beta/2N+i)^{4N}}
\lim_{u\to 0}
\frac{1}{4u}\left[(2u+i)
((u-i \beta/2N+i)(u+i \beta/2N))^{2N}
\times
\prod_{k=1}^{2n}
\frac{(u-\tilde\lambda_k-i)}{(u-\tilde\lambda_k)}\right.
\\
&\left.
+(2u-i)
((u+i \beta/2N-i)(u-i \beta/2N))^{2N}
\times
\prod_{k=1}^{2n} \frac{(u-\tilde\lambda_k+i)}{(u-\tilde\lambda_k)}
\right].
\end{split}
\end{equation}

We performed exact diagonalization of $\mathcal{T}$ for small values of $N$ and found that at
the
largest eigenvalue is given by the unique state with $N$ purely real
roots. Similar to the XXZ case we define the auxiliary function
\begin{equation*}
  \fa(u)=
K_D(u)
\left[
\frac{(u+i \beta/2N-i)}{(u-i \beta/2N+i)}
\frac{(u-i \beta/2N)}{(u+i \beta/2N)}
\right]^{2N}
\prod_{k=1}^{2N}
\frac{(u-\tilde\lambda_k+i)}{(u-\tilde\lambda_k-i)}
\end{equation*}
and 
\begin{equation*}
  A(u)=\frac{1+\fa(u)}{1+K_D(u)}.
\end{equation*}
Also, we define the canonical contour $C$ which now consists of two infinite
horizontal lines with imaginary parts $\pm\alpha$ such that $\alpha<
1/2$. The function $A(u)$ has zeroes inside the contour given by the
(doubled set of) Bethe roots. The
trivial zero of $1+\fa(u)$ at $u=0$ is canceled by the denominator,
and further numerical checks showed that for the leading state there are indeed no
other zeroes inside $C$. 

It is now a straightforward exercise to derive an NLIE for the
auxiliary function. Here we just give the result valid in the
Trotter limit:
\begin{equation}
\label{BNLIE-XXX}
\begin{split}
\log \fa(\lambda)=&
\log(K_D(u))+s \frac{4}{\lambda(\lambda+i)}
-  \int_C \frac{d\omega}{2\pi } 
\frac{2}{(\lambda-\omega)^2+1}  \log(A(\omega)).
\end{split}
\end{equation}
Expressing the eigenvalue as an integral is more
involved because \eqref{LaD} is not of a product form. 
However, the same manipulations which are used in the periodic case \cite{kluemper-review}
 can be performed here as well and in the Trotter limit we find 
\begin{equation}
  \label{g-qtm-XXX}
  g(s)=\lim_{N\to\infty} \frac{\log \Lambda}{2}=
-\frac{1}{2}\int_{C}\frac{d\omega}{2\pi } 
\frac{ \log(A(\omega))}{\omega(\omega+i)}.
\end{equation}
Equations \eqref{BNLIE-XXX}-\eqref{g-qtm-XXX} are the main results
of this section. 

For the sake of completeness we note that in the case of the Neel state 
the same equations apply with the only difference that the reflection
factor $K_D(u)$ has to be replaced by
\begin{equation*}
  K_N(u)=\frac{(u+i)(2u-i)}{(u-i)(2u+i)}.
\end{equation*}

\subsection{The $s=0$ limit}

At $s=0$ the solution of the NLIE \eqref{BNLIE-XXX} is simply $\fa(u)=K_D(u)$ and this
yields $g(0)=0$ as expected. As a non-trivial check we compute the
expectation value of the Hamiltonian from the NLIE:
\begin{equation*}
\frac{\bra{D}H\ket{D}}{L}=
-\frac{\partial g}{\partial s}=
-\int_C\frac{d\omega}{2\pi }
\frac{1}{\omega(\omega+i)}
 \frac{\fa'(\omega)}{1+K_D(-\omega)}.
\end{equation*}
Here $\fa'(u)$ is the solution of the linear integral equation
\begin{equation*}
\begin{split}
  \fa'(u)=
 \frac{4}{u(u+i)}
-\int_C\frac{d\omega}{2\pi}
\frac{2}{(u-\omega)^2+1}
\frac{\fa'(\omega)}{1+K_D(-\omega)}.
\end{split}
\end{equation*}
The solution is
\begin{equation*}
  \fa'(u)=
4i\frac{u^2-1}{u(u^2+1)}.
\end{equation*}
This leads to
\begin{equation*}
  \frac{\bra{D}H\ket{D}}{L}=
-\int_C\frac{du}{2\pi}
\frac{1}{u(u+i)}
2i\frac{u^2-1}{u(u^2+1)}
\frac{2u-i}{u}=-\frac{5}{2}.
\end{equation*}
It can be checked by a straightforward real-space calculation that
this is indeed the correct expectation value. 

\subsection{The $q$-deformed dimer state}

The natural generalization of the dimer state to $\Delta\ne 1$ is the so-called $q$-deformed dimer state:
\begin{equation}
\label{qD}
  \ket{qD}=\otimes_{j=1}^{L/2} \left(\frac{\ket{+-}-q\ket{-+}}{\sqrt{1+q^2}}\right),
\end{equation}
where $\Delta=(q+1/q)/2$. This state (together with its translation by a site one) is the ground state of the $q$-deformed Majumdar-Ghosh Hamiltonian derived in \cite{q-mg}.

Consider the function $g(s)$ defined as
\begin{equation*}
  e^{g(s)L}=\bra{qD} e^{-sH}\ket{qD},
\end{equation*}
where the anisotropy $\Delta$ of the Hamiltonian is the same as that of the initial state. Based on the previous calculations it is easy to see that equations \eqref{BQTMhoho}-\eqref{BQTMhehe}  hold also in this case with
\begin{equation*}
  K(u)=\frac{\sinh(2u-\eta)}{\sinh(2u+\eta)}.
\end{equation*}

\section{Conclusions}

In this work we derived exact analytical results for the dynamical
free energy density (the Loschmidt echo for imaginary times) for
certain quantum quenches in the XXZ spin chain. As initial states we
considered the N\'eel state and the ($q$-deformed) dimer state, which
are both products of local two-site states. In all cases considered
the resulting equation for the dynamical free energy takes the form 
\begin{equation}
\label{BQTMhehe2}
  g(s)=
\frac{1}{2}\int_C\frac{d\omega}{2\pi i}
\tilde e(u)
 \log \left(\frac{1+\fa(u)}{1+K(u)}\right),
\end{equation}
where $\fa(u)$ is the solution of the NLIE
\begin{equation}
\label{BQTMhoho2}
\begin{split}
  \log \fa(u)=
\log(K(u))
-2s e(u)
-\int_C\frac{d\omega}{2\pi i}
\varphi(u-\omega)
 \log \left(\frac{1+\fa(u)}{1+K(u)}\right).
\end{split}
\end{equation}
The function $e(u)$ is a function related to the one-particle energy (differing
from $\tilde e(u)$ by a simple proportionality factor), $\varphi(u)$
is the one-particle scattering kernel, and the complex contour $C$
depends on the anisotropy $\Delta$. The difference between these
equations and those describing the thermodynamic free energy density
\cite{kluemper-review} are the appearance of the extra source term
$\log(K(u))$ (which carries the information about the initial state)
and the regulator $1/(1+K(u))$ for the integrals which was introduced
to correctly handle the zeroes of the function $1+\fa(u)$. 

We observed that the NLIE is numerically stable in a finite
neighbourhood of the real positive axis including the purely imaginary
values $s=it$ with small $t\in\valos$, therefore it is capable of
providing exact results for the Loschmidt echo per site. We found that
as $t$ is increased the zeroes of the function $1+\fa(u)$ move
outwards from the origin, and increasing $t$ further they would
eventually cross the
canonical contour. In this case the equations have to be modified
accordingly. However, the NLIE becomes numerically unstable
even before the first crossing appears, and further work is needed to
obtain numerical data for larger real times. Also, it needs to be
checked whether and in what cases level crossings of the Boundary Quantum Transfer Matrix
happen at real times, which 
could lead to non-analyticity of the Loschmidt echo. We plan to return
to these questions in a future work.

The reason for choosing the N\'eel state and the ($q$-deformed) dimer
state was that in these cases the additional zero of the function
$1+\fa(u)$ within the canonical contour which does not correspond to a Bethe
root is fixed to $u=i\pi/2$ or $u=0$. In the
 case of
\begin{equation}
\label{aaa}
\ket{  \Psi_0}=\otimes_{j=1}^{L/2}\ \frac{\ket{+-}+\gamma\ket{-+}}{1+|\gamma|^2}
\end{equation}
with arbitrary $\gamma$ the position of the additional zero depends
both on $s$ and $\gamma$. In this case a
different NLIE can be written down which accounts for the movement of
the 
additional zero.

More general two-site states could be considered by taking
off-diagonal $K$-matrices for the construction of the boundary
transfer matrix \eqref{boundaryT}. The
diagonalization of these transfer matrices could  be achieved using
the recent results of \cite{off-cao,off-nepo}.

It is an intriguing question whether generalizations of the present
methods could lead to analytic expressions for the time-dependent
local correlation functions, possibly through a limit 
\begin{equation}
\label{oooo}
\ordo(t)=\lim_{\beta\to 0}
\bra{\Psi_0} e^{(it-\beta) H} \ordo e^{(-it-\beta) H}\ket{\Psi_0}.
\end{equation}
Results for dynamical correlation functions at equilibrium are already
available with the QTM method \cite{QTM-real-time}, but it is not
evident  
if such methods could work for the Boundary QTM relevant to the quench problem.
While it is straightforward to translate the r.h.s. of \eqref{oooo} to
a 6-vertex model partition function and the leading state of the
relevant Boundary Quantum Transfer Matrix could be constructed using
the methods of the present work, the insertion of the local operators
leads to scalar products whose evaluation is very challenging even at
finite Trotter number.  

\bigskip

While this work was being finished, the paper \cite{Fagotti-Loschmidt}
appeared where M. Fagotti  derived a different NLIE for $g(s)$
using the GGE hypothesis for the
exponential of the Hamiltonian. Whereas the source term of the NLIE of
\cite{Fagotti-Loschmidt} is not explicit (the Lagrange-multipliers are
not specified) analytical or numerical results could still be obtained
for small quenches, for example using a $1/\Delta$ expansion for the
quench starting from the N\'eel state. Also, it was observed in
\cite{Fagotti-Loschmidt} that the winding number of the auxiliary
function changes as the time $t$ is increased.  This is in complete
accordance with the movement of the zeroes of $A(u)$ observed in the
present work.

We believe that it is worthwhile to stress the differences between our
results and those of \cite{Fagotti-Loschmidt}. Our lattice path
integral derivation of $g(s)$ is based on first principles, the
assumptions about the analyticity of the auxiliary functions were
checked numerically at finite and infinite Trotter number, and we did not make use
of the GGE hypothesis. Therefore our NLIE gives exact
result for $g(s)$ for any $s\in\valos^+$ and 
 both the source term $\log(K(u))$ and
the regulator $1/(1+K(u))$ are explicit.  
On the other hand, the
analytic continuation to real times ($s=it$) and in general the
behaviour of $g(s)$ for complex $s$ deserves further study. 
Also, it is an interesting question whether there is a direct
relation between our NLIE and that of
\cite{Fagotti-Loschmidt}. Comparing our numerical results to those of
\cite{Fagotti-Loschmidt} could lead to a first  check of the
GGE hypothesis for the XXZ spin chain.
These
questions are left for future research. 

\bigskip
\textbf{Acknowledgements}
\medskip

We are grateful to G\'abor Tak\'acs for motivating discussions and
useful comments on the manuscript.

This research was started while the author was employed by the NWO/VENI
grant 016.119.023 at the University of Amsterdam, the Netherlands. 

The second half of the work was realized
in the frames of TAMOP 4.2.4. A/1-11-1-2012-0001
 ,,National Excellence Program -- Elaborating and operating an inland
 student and researcher personal support system convergence program''.
 The project was subsidized 
by the European Union and co-financed by the European Social Fund.

\addcontentsline{toc}{section}{References}
\bibliography{BQTM-gyors}
\bibliographystyle{utphys}

\end{document}